\documentclass[aps,prl,10pt,amsmath,floats,floatfix,twocolumn,superscriptaddress,altaffilletter,nofootinbib,shownopacs,longbibliography,frontmatterverbose]{revtex4-1}
\newcommand\prlsec[1]{\vspace{2mm} \textbf{\emph{#1}}\,---}
\usepackage[T1]{fontenc}
\usepackage[utf8]{inputenc}
\usepackage{textcomp}
\usepackage{lmodern}
\usepackage{verbatim}
\usepackage{physics}
\usepackage[dvipsnames]{xcolor}
\definecolor{linkcolor}{rgb}{0.6,0.0,0.0}
\usepackage{appendix}
\usepackage{booktabs}
\usepackage{enumitem}
\usepackage[hypertexnames=false, unicode, colorlinks=true, linkcolor=linkcolor,
citecolor=linkcolor, filecolor=linkcolor,urlcolor=linkcolor,
pdfusetitle]{hyperref}
\usepackage[all]{hypcap}
\usepackage{graphicx}
\usepackage{color}
\usepackage{orcidlink}
\usepackage{xspace}
\usepackage{amssymb}
\usepackage[normalem]{ulem}
\usepackage{bm}
\usepackage{orcidlink}
\usepackage{microtype}
\usepackage[english]{babel}
\usepackage{blindtext}
\usepackage{scrextend}
\usepackage{hyperref}
\usepackage{lineno}
\newcommand{\du}{\mathrm{d}}

\newcommand{\eventname}{GW241011\xspace}
\newcommand{\evcsq}{\ensuremath{\mathrm{eV} \, c^{-2}\xspace}}

\newcommand{\fulltitle}{Implications of \eventname  for rotating exotic compact objects}

\footnotetext{These authors contributed equally to this work.
}

\begin{document}
\title{\fulltitle}
\author{N. V. Krishnendu\orcidlink{0000-0002-3483-7517}\textsuperscript{*}}
\email{k.naderivarium@bham.ac.uk}
\affiliation{Institute for Gravitational Wave Astronomy \& School of Physics and Astronomy, University of Birmingham, Birmingham, B15 2TT, United Kingdom}
\author{Tamara Evstafyeva\orcidlink{0000-0002-2818-701X}\textsuperscript{*}}
\email{tevstafyeva@perimeterinstitute.ca}
\affiliation{Perimeter Institute for Theoretical Physics, Waterloo, Ontario N2L 2Y5, Canada}
\author{Aditya Vijaykumar\orcidlink{0000-0002-4103-0666}\textsuperscript{*}}
\email{aditya@utoronto.ca}
\affiliation{Canadian Institute for Theoretical Astrophysics, University of Toronto, 60 St George St, Toronto, ON M5S 3H8, Canada}
\author{William E. East\orcidlink{0000-0003-0988-4423}}
\email{weast@perimeterinstitute.ca}
\affiliation{Perimeter Institute for Theoretical Physics, Waterloo, Ontario N2L 2Y5, Canada}
\author{Rimo Das\orcidlink{0009-0002-8388-0922}}
\affiliation{Department of Physics, Indian Institute of Technology Madras, Chennai 600036, India}

\author{Sayantani Datta}
\affiliation{Department of Physics, University of Virginia, P.O. Box 400714, Charlottesville, Virginia 22904-4714, USA}
\affiliation{Inter-University Centre for Astronomy and Astrophysics, PostBag 4, Ganeshkhind, Pune-411007, India}
\author{Nils Siemonsen\orcidlink{0000-0001-5664-3521}}
\affiliation{Princeton Gravity Initiative, Princeton University, Princeton, New Jersey 08544, USA}
\affiliation{Department of Physics, Princeton University, Princeton, New Jersey 08544, USA}
\author{Nami Uchikata}
\affiliation{Institute for Cosmic Ray Research, KAGRA Observatory, The University of Tokyo, 5-1-5 Kashiwa-no-Ha, Kashiwa City, Chiba 277-8582, Japan}
\author{Poulami Dutta Roy\orcidlink{0000-0001-8874-4888}}
\affiliation{Department of Physics, University of Florida, PO Box 118440, Gainesville, Florida 32611-8440, USA.}

\author{Anuradha Gupta\orcidlink{0000-0002-5441-9013}}
\affiliation{Department of Physics and Astronomy, University of Mississippi, University, Mississippi 38677, USA}

\author{Ish Gupta\orcidlink{0000-0002-1654-7348}}
\affiliation{Department of Physics, University of California, Berkeley, CA 94720, USA}
\affiliation{Department of Physics and Astronomy, Northwestern University, 2145 Sheridan Road, Evanston, IL 60208, USA}
\affiliation{Center for Interdisciplinary Exploration and Research in Astrophysics (CIERA), Northwestern University, 1800 Sherman Ave, Evanston, IL 60201, USA}

\author{Syed U. Naqvi\orcidlink{0000-0002-9380-0773}} 
\affiliation{Institute of Nuclear Physics, Polish Academy of Sciences, Radzikowskiego 152, 31-342 Kraków, Poland}

\author{Manuel Piarulli\orcidlink{0009-0009-4099-9166}}
\affiliation{Laboratoire des 2 Infinis - Toulouse (L2IT-IN2P3), Université de Toulouse, CNRS, F-31062 Toulouse Cedex 9, France}

\author{Muhammed Saleem\orcidlink{0000-0002-3836-7751}}
\affiliation{Center for Gravitational Physics, The University of Texas at Austin, Austin, Texas 78712, USA}

\author{Elise M.~S\"anger\orcidlink{0009-0003-6642-8974}}
\affiliation{Max Planck Institute for Gravitational Physics (Albert Einstein Institute), Am M{\"u}hlenberg 1, Potsdam, 14476, Germany}
\author{Pratyusava Baral\orcidlink{0000-0001-6308-211X}}
\affiliation{Department of Physics \& Astronomy, University of Wisconsin-Milwaukee, Milwaukee, WI 53211}
\author{Sajad A. Bhat\orcidlink{0000-0002-6783-1840}}
\affiliation{Inter-University Centre for Astronomy and Astrophysics, PostBag 4, Ganeshkhind, Pune-411007, India}
\author{Thomas A.~Callister\orcidlink{0000-0001-9892-177X}}
\affiliation{Williams College, Williamstown, MA 01267, USA}

\author{Mattia Emma\orcidlink{0000-0001-7943-0262}}
\affiliation{Department of Physics, Royal Holloway University of London, Egham, TW20 0EX}

\author{Carl-Johan Haster\orcidlink{0000-0001-8040-9807}}
\affiliation{Department of Physics and Astronomy, University of Nevada, Las Vegas, NV 89154, USA}
\affiliation{Nevada Center for Astrophysics, University of Nevada, Las Vegas, NV 89154, USA}

\begin{abstract}
A number of theoretical proposals have been made for horizonless compact objects with masses and spins similar to those of black holes.
While gravitational wave signatures from their mergers can resemble those of black holes, features like the spin-induced quadrupole moment may reveal their distinct nature. 
Using the tight bounds on the spin-induced quadrupole moment of \eventname, we place gravitational wave constraints on the nature of its primary. 
We find that large classes of exotic compact objects (including rotating boson stars) cannot explain its nature, however, models of sufficiently large compactness of $C \gtrsim 0.24$ may still be viable contenders. 
\end{abstract}
\maketitle
\prlsec{Introduction}
The latest LIGO-Virgo-KAGRA (LVK)~\cite{LIGOScientific:2014pky, VIRGO:2014yos, KAGRA:2020tym} gravitational wave (GW) transient catalog reported over 200 compact binary coalescences~\cite{LIGOScientific:2025yae}. 
Owing to increased sensitivity and several exceptional events~\cite{LIGOScientific:2024elc,LIGOScientific:2025cmm,GW231123,GW250114_LVK_Area_Law,GW250114:LVK_follow_up, gw241011,LIGOScientific:2025slb}, we are in the position to
perform unprecedented tests of general relativity and the Standard Model of particle physics~\cite{GWTC-3-TGR}. One such exemplary event is \eventname~\cite{gw241011,ligo_scientific_collaboration_2025_17343574}, a compact binary
with a highly spinning primary (dimensionless spin $\chi_1 \sim 0.78_{-0.09}^{+0.09}$) that is roughly three times more massive than the secondary (primary mass $m_1 \sim 19.6_{-2.5}^{+3.6} M_{\odot}$ and mass ratio $q \sim 0.30_{-0.08}^{+0.09}$).
These properties, together with its high signal strength,
 make \eventname a perfect laboratory for investigating the fundamental question at the center of this work: do compact objects beyond the realm of black holes (BHs), neutron stars (NSs), and white dwarfs populate our Universe?

There have been a variety of proposed models for such
exotic compact objects (ECOs)~\cite{Cardoso:2019rvt,Bezares:2024btu,Sennett:2017etc, Cardoso:2017cfl}. Among the most studied are scalar and vector boson stars (BSs)~\cite{Liebling2012,Kaup:1968zz}, while other proposals, such as gravastars (GSs)~\cite{Mazur:2004fk,Mazur:2001fv,Posada:2018agb,Pani:2009ss,Cardoso:2016oxy, Visser:2003ge} and anti-de Sitter (AdS) black shells~\cite{Danielsson:2017riq, Giri:2024cks}, are still being developed. While the GW phenomenology of BSs has been explored through numerical relativity in some regions of the parameter space~\cite{Evstafyeva:2022bpr, Evstafyeva:2024qvp,Siemonsen:2023hko,Bezares:2022obu,Sanchis-Gual:2022mkk}, 
much of the landscape of ECO signatures still awaits a systematic study. 
Given the present challenges associated with accurately modeling GW signatures of ECOs, agnostic parameterized tests of general finite-size signatures~\cite{Johnson-Mcdaniel:2018cdu,Ghosh:2025wex,Krishnendu:2017shb, Cardoso:2019rvt,Chia:2023tle, Krishnendu:2019tjp, KY19, GW190814, GW190425}, i.e., effects related to the body’s internal structure, currently provide the most robust approach to probing the nature of these compact objects.
 
Spin-induced multipole moments are one class of finite-size effects. They arise when the spin of a compact object deforms its shape and generates a hierarchy of multipoles that imprint characteristic signatures on the emitted GW signal. While these multipoles are uniquely determined for Kerr BHs by their mass and spin~\cite{Poisson:1997ha,Th80, Hansen:1974zz, Carter71,Laar97}, non-BH compact objects can have spin-induced deformations  distinct from that of BHs~\cite{Pacilio:2020jza,Ryan97, Vaglio:2022flq,Sukhov:2024bwo,Uchikata2016,Uchikata:2021jmy, Ryan97b,Cardoso:2019rvt}, allowing for tests of the Kerr geometry 
through the GW signal of compact binaries. 
In particular, the leading-order spin-induced multipole---the spin-induced quadrupole moment (SIQM)---can be used to constrain the NS equation of state (EOS)~\cite{Harry:2018hke}, or various models of ECOs~\cite{Vaglio:2023lrd}. 
The SIQM (and higher-order moments) of \textit{spinning} ECOs is less well-studied than that of NSs. 
Current efforts have focused on computing SIQM for certain BS models in the large self-interaction approximation~\cite{Ryan97b,Vaglio:2022flq} and from numerically computed spacetimes~\cite{Adam:2022nlq,Adam:2024zqr,Sukhov:2024bwo}; as well as slowly rotating thin-shell GSs~\cite{Uchikata2015,Uchikata2016}.
\begin{figure}[!htbp]
    \includegraphics[width= 3.5 in]{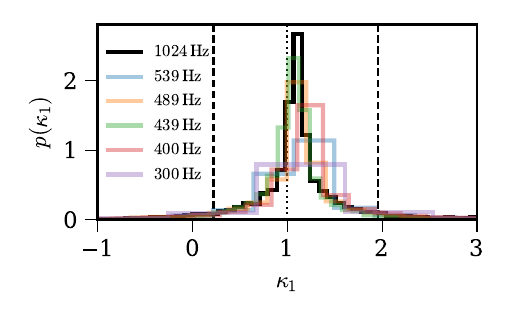}
    \caption{The one-dimensional posterior distribution of $\kappa_1$ obtained from \eventname.
    The colored lines represent posteriors obtained from the likelihood truncated at different frequencies $f_{\rm cut}$,
    whereas the black line corresponds to the full inference. The dashed vertical lines represent the 95\% CI bounds and the dotted vertical line denotes the corresponding Kerr BH value of $\kappa_1 = 1$.
    } 
    \label{fig:K1_1D}
\end{figure}

Remarkably, \eventname~\cite{gw241011} placed the most stringent constraint on beyond-Kerr corrections to the SIQM of the primary to date~\cite{gw241011}. Here, we utilize this measurement to constrain the parameter space of ECO models that are still compatible with the primary of \eventname. 
In particular, we compute the SIQMs of spinning BSs and place constraints on their potential and the scalar mass. We also extend our constraints to exotic fluid stars (EFSs) with sound speed equal to the speed of light, as a way to quantify the lower bound on the compactnesses of exotic stars that are not yet ruled out by \eventname.
Finally, we comment on the implications for existing spinning gravastar models, dynamically scalarized
BHs, and Proca stars, as well as the limitations of current modeling assumptions.

Unless otherwise stated, we use geometric units with $G=c=1$.

\prlsec{SIQM measurements from \eventname } 
The post-Newtonian (PN) formalism provides a framework to describe the inspiral dynamics of a spinning compact binary~\cite{Bliving,Marsat:2014xea,Bohe:2015ana}. Within this formalism, leading-order SIQM corrections to the GW phase appear at 2PN order~\cite{Mishra:2016whh, Poisson:1997ha}. The SIQM of each compact object scales as $Q_i =-\kappa_i \chi_i^2 m_i^3$~\cite{Hansen:1974zz, Carter71, Poisson:1997ha}, where $i \in \{1,2 \}$, corresponding to the more and less massive object, respectively, and $\kappa_i$ denotes the \textit{reduced} quadrupole. For Kerr BHs, $\kappa_i=1$, but alternative objects can have different values~\cite{Pacilio:2020jza,Ryan97, Vaglio:2022flq,Sukhov:2024bwo,Uchikata2016,Uchikata:2021jmy}. 
By explicitly introducing parametric coefficients $\kappa_i$ in the inspiral phase, we can therefore test the binary progenitors for deviations from the Kerr hypothesis.
We use the implementation of the SIQM corrections at 2PN and 3PN orders developed in Ref.~\cite{Divyajyoti:2023izl} within the \textsc{IMRPhenomXPHM} waveform model~\cite{Pratten:2020ceb}, which includes double-spin precession, higher harmonics, and a complete inspiral–merger–ringdown treatment for binary black holes. As the waveform model does not readily support super-extremal spins $\chi > 1$ predicted by some BS and GS configurations, we restrict our discussion to sub-extremal spins, i.e.,~$\chi_{1,2} \in [0,0.99]$.

In Fig.~\ref{fig:K1_1D}, we illustrate the $\kappa_1$ posterior probability distributions
from \eventname~\cite{gw241011} and highlight the $95\%$ credible interval (CI) with black dashed lines. The curves in the background are obtained by varying the maximum cut-off frequency $f_{\rm{cut}}$ when evaluating the likelihood function. For lower values of $f_{\rm cut}$, posteriors on $\kappa_1$ are wider as a result of the reduced signal. The qualitative behavior of the posteriors remains unchanged across different $f_{\rm{cut}}$, highlighting the robustness of the $\kappa$ constraints (see~Supplementary Material for more details). The SIQM of the secondary, $\kappa_2$, on the other hand, is not constrained away from its prior; we therefore do not discuss it here. 
We utilize the SIQM measurement from \eventname to put stringent constraints on ECO models. 
For a broad range of ECOs, we identify models that are consistent with the spin constraints of the primary in \eventname and then compare the inferred SIQM measurement with their theoretical predictions. As such, our constraints only apply to the primary object in \eventname and do \textit{not} rule out the existence of ECOs in other systems. In what follows, our comparisons use the full two-dimensional 95\% isocontours of the $\kappa_1 - \chi_1$ posterior of the primary object in \eventname. Unless otherwise stated, all ECO parameters we constrain are quoted within the 95\% CI.

\begin{figure}[!htbp]
    \includegraphics{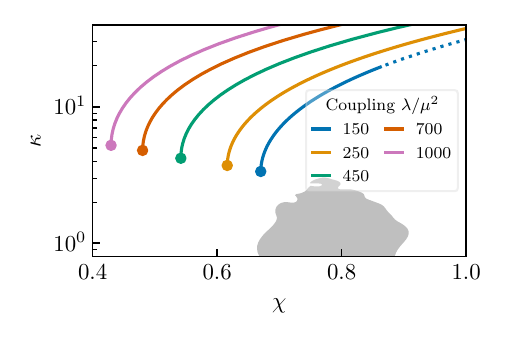}
    \includegraphics{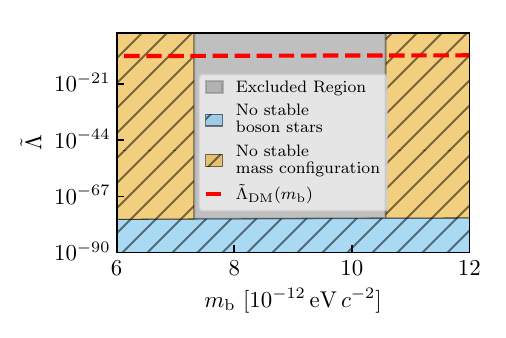}
    \caption{
    \textit{Top:} Sequences of repulsive BSs in the $\chi-\kappa$ plane (solid lines), with 
    different colors denoting the coupling values, $\lambda / \mu^2$, and dots the maximum mass solution. The dotted lines indicate dynamically unstable solutions according to our relativistic criterion (see the main text). 
    The gray shaded region denotes the $\kappa$--$\chi$ measurement (at 95\% credible level).
    \textit{Bottom:} Exclusion regions in the $m_\mathrm{b} - \tilde{\Lambda}$ plane, where $m_\mathrm{b}$ is the boson mass in physical units, $\tilde{\Lambda} = \Lambda (\hbar c)$ is dimensionless and $\Lambda$ is the coupling in physical units. Gray denotes the region excluded by \eventname, 
     yellow corresponds to stars with masses incompatible with \eventname, whereas blue contains only unstable solutions.
    For reference, we also overplot the $\tilde{\Lambda}$ and $m_{\rm b}$ values corresponding to a hypothetical dark matter particle with cross-section per unit mass of $0.1 \,\mathrm{cm}^{2}\,\mathrm{g}^{-1}$ in dashed red line. 
    }
    \label{fig:repulsive}
\end{figure}

\prlsec{Boson stars} 
Scalar boson stars arise from a massive complex scalar field $\varphi$, characterized by a potential $V(|\varphi|)$, that is minimally coupled to gravity. 
We adopt the three commonly studied potentials: (i) repulsive, $V_{\rm{rep}}(|\varphi|^2) = \mu^2 |\varphi|^2 + \lambda |\varphi|^4$; (ii) solitonic, $V_{\rm{sol}}(|\varphi|^2) = \mu^2 |\varphi|^2 \left(1 - \frac{2|\varphi|^2}{\sigma^2} \right)^2$; and (iii) axionic, $V_{\rm axion}(|\varphi|^2) = \mu^2 f^2 \left\{ 1 - \cos \left[ \sqrt{2 |\varphi|^2} f^{-1} \right] \right\}$. 
Here, $\mu$ denotes the mass parameter, whilst $\lambda/\mu^2$, $\sigma$, and $f$ are dimensionless coupling constants that quantify self-interactions. Our $\mu$ relates to the boson mass $m_{\rm b}$ via $\mu = m_{\rm b} c/\hbar$.
Spinning BSs are modeled by the scalar ansatz $\varphi = A(r, \theta) e^{i(\omega_{\rm{BS}} t + m \phi)}$, where $A$ denotes the amplitude, $\omega_{\rm{BS}}$ the BS frequency, and $m$ the azimuthal number. 

For a fixed potential and coupling, we compute a family of BS solutions and their SIQMs (cf.~Supplementary Material for more details). However, not all configurations we construct are viable candidates for ECOs due to stability considerations. We therefore exclude unstable stars, as indicated by (i) catastrophe theory turning-point arguments~\cite{Kleihaus:2011sx}, and (ii) numerical simulations, identifying linear non-axisymmetric instabilities in models with no or small self-interactions~\cite{Sanchis-Gual:2019ljs,Siemonsen:2020hcg}. Additionally, we explicitly set $m=1$ in the scalar field ansatz, as no dynamically stable boson star configurations with $m>1$ have been obtained yet. 

For item (ii), we determine the regions of dynamical stability using the results of numerical simulations of Ref.~\cite{Siemonsen:2020hcg}. For instance, for repulsive stars, we extract the minimum threshold on the coupling to be $\lambda > 210/M_{\rm{BS}}^2$, where $M_{\rm{BS}}$ is the BS mass\footnote{A similar estimate of $\lambda > 168/M_{\rm{BS}}^2$ has been reported in Ref.~\cite{Dmitriev:2021utv} for (Newtonian) Bose stars, which is in remarkably good agreement with our relativistic threshold.}. 
For solitonic and axionic stars, we determine dynamical stability by the critical BS frequency $\omega_{\rm{BS},c}$: the stars tend to stabilize at lower frequencies $\omega_{\rm{BS}} < \omega_{\rm{BS,c}}$, corresponding to more compact solutions (see, e.g., Fig.~5 of Ref.~\cite{Siemonsen:2020hcg}). 
For solitonic models, we expect models consistent with \eventname to be dynamically stable for $\sigma \lesssim 0.1$, whilst for axionic stars we do not have enough simulation data to draw concrete conclusions. 

\textit{Repulsive stars:} We first start our discussion on constraints with the repulsive potential. In the top panel of Fig.~\ref{fig:repulsive}, we illustrate our numerically constructed families of repulsive BSs for different couplings $\lambda/\mu^2 \in [150, 1000]$ in the $\kappa$--$\chi$ space.
All models (apart from solutions with $\lambda /\mu^2=150$ and $\chi > 0.86$) consistent with the $\chi_1$ 
measurement of \eventname permit dynamically stable solutions according to our relativistic stability criterion. If a repulsive potential is to remain consistent with the primary mass and spin measurements of \eventname, we find that the boson mass would lie within the range $m_{\rm b} \in [7.32$\textendash$ 10.6] \times 10^{-12}\, \evcsq$. 
However, the smallest theoretically-allowed $\kappa$ consistent with the primary spin of \eventname is ${\sim} 3$ (corresponding to a smaller coupling $\lambda / \mu^2 = 150$)---much larger than the limits derived from \eventname. We thus conclude that the primary of \eventname cannot be explained by a repulsive BS, and exclude $\lambda / \mu^2 \geq 150$ for the aforementioned $\mu$ range. For very large couplings $\lambda / \mu^2 \gtrsim 450$, solutions compatible with the spin of \eventname have larger radii and achieve maximum compactness of $C_{\rm max} \leq 0.08$. Naturally, these models are therefore unlikely to explain the nature of \eventname. However, for $\lambda / \mu^2 \leq 250$, we find $C_{\rm max} \sim 0.11 - 0.17$, corresponding to more slowly spinning models. 
In the Supplementary Material, we show that there is significant coherent power in the signal past the nominal cut-off frequencies for these compactness values, further suggesting that such low compactness values are ruled out.

In the bottom panel of Fig.~\ref{fig:repulsive}, we additionally map the constraints on the repulsive potential onto the exclusion region in $m_{\rm b}$\textendash$\tilde{\Lambda}$ plane, where $\Lambda$ is the coupling in physical units and $\tilde{\Lambda} = \Lambda (\hbar c)$ is
dimensionless. 
We then compare it to 
the upper limits on the cross-section of self-interacting dark matter $\sigma^{\rm max}_{\rm cross}/m_{\rm b} \approx 0.1 \, \mathrm{cm}^2 \, \mathrm{g}^{-1}$ from various observations (see, e.g., Refs.~\cite{Harvey:2018uwf, Correa:2020qam, Andrade:2020lqq}). Using the tree-level expression for the cross section per unit mass~\cite{Choi:2016hid}, $\sigma_{\rm cross} / m_{\rm b} = 3\, (32 \pi m_\mathrm{b})^{-1} (\Lambda \hbar^2 / m_\mathrm{b})^2$,
we find that couplings corresponding to cross sections below $\sigma^{\rm{max}}_{\rm cross} / m_{\rm b}$
are excluded by the data within the boson mass range probed by \eventname.  We stress again that these exclusion regions \textit{only} apply to the primary in \eventname and cannot be extended to other objects in the Universe.

\begin{figure}[!thbp]
    \centering
    \includegraphics{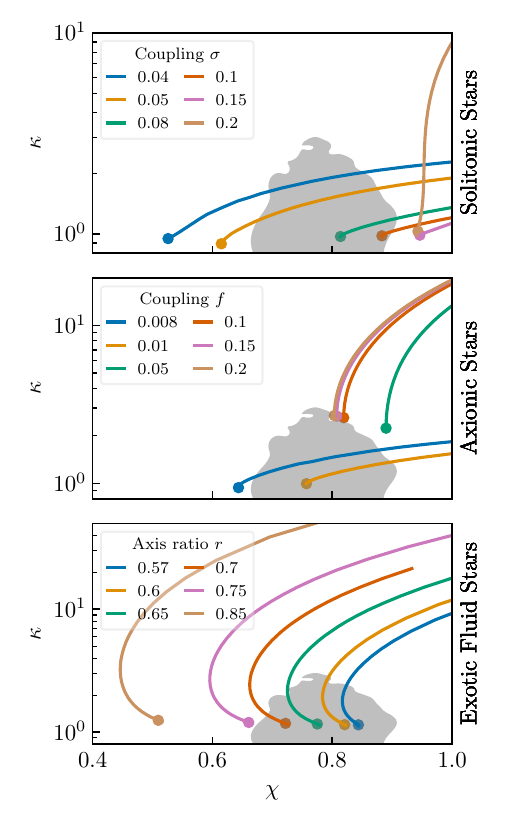}
    \caption{Same as Fig.~\ref{fig:repulsive}, but for 
    solitonic stars (top), axionic stars (middle), and exotic fluid stars (bottom), with colors corresponding to values of $\sigma$, $f$, and $r$ respectively\footnote{For solitonic $\sigma=0.04$ models the dot corresponds to the end of the numerically computed sequence.}. Here, however, we do not indicate solutions that suffer from dynamical instabilities.
    }
    \label{fig:solitonic-axionic-fluid}
\end{figure}
\textit{Solitonic stars:} The constraints are more subtle for the solitonic potential, as illustrated in the top panel of Fig.~\ref{fig:solitonic-axionic-fluid}. First, for smaller self-interactions ($\sigma > 0.1$), there are no  solutions consistent with $\chi_1$ of \eventname (these solutions are also likely to be dynamically unstable). Larger self-interactions ($\sigma < 0.1$), on the other hand, do permit stable solutions that are consistent with both $\chi_1$ and $\kappa_1 $ measurements. The relatively low values of $\kappa$ here arise from higher compactness ($C \gtrsim 0.26$) of these solitonic stars, in contrast to repulsive models. The allowed gray region in fact even intersects with some ultracompact stars with light rings~\cite{Evstafyeva:2025mvx}. 
Apart from some models with $\sigma = 0.05$ and $\sigma = 0.04$ at high spins, we conclude that, overall, solitonic stars with large self-interactions $\sigma_0 \lesssim 0.1$ and boson masses $m_{\rm b} \in [1.4$\textendash$4.5] \times 10^{-11} \, \evcsq$ are generally consistent with \eventname.

\textit{Axionic stars: } Similar to the solitonic case, large self-interactions of $f \leq 0.01$ in the axionic potential also lead to highly compact models ($C \sim 0.3$) that are consistent with the spin region of \eventname, and therefore the $\kappa_1$ measurement (see the middle panel of Fig.~\ref{fig:solitonic-axionic-fluid}). These models have $m_{\rm b} \in [2.6$\textendash$5.0]\times10^{-11}\, \evcsq$. On the other hand, for smaller self-interactions of $f \gtrsim  0.05$, axionic models compatible with \eventname are ruled out; these models have $C \leq 0.17$ and $m_{\rm b} \in [7.1$\textendash$9.5]\times10^{-12}\, \evcsq$. However, solutions with $f>0.1$ are more likely to suffer from dynamical instabilities~\cite{Siemonsen:2020hcg}. 

\prlsec{Exotic fluid stars}As a toy model for ECOs, we additionally consider 
EFSs with a compressible EOS at the causal limit of stiffness (i.e.,~the sound speed equals the speed of light). 
The pressure is
\begin{equation}
P = 
\begin{cases}
    \begin{aligned}
    (\epsilon - \epsilon_s), \quad &\epsilon \geq \epsilon_s, \\
    0, \quad &\epsilon < \epsilon_s,
    \end{aligned}
\end{cases}
\end{equation}
where $\epsilon$ is the total energy density and $\epsilon_s$ some critical energy density, setting the scale of the problem. We construct solutions with uniform rotation by varying the ratio of polar to equatorial radius, i.e., the axis ratio $r$, using the \textsc{rns} package~\cite{Nozawa:1998ak,Stergioulas:1994ea,Cook1992:amns} (see~Supplementary Material). 
Changing $\epsilon_s$ corresponds to a simple rescaling of dimensionful quantities, e.g., mass~$M_{\rm{EFS}} \propto (\sqrt{\epsilon_s})^{-1}$, radius~$R_{\rm{EFS}} \propto (\sqrt{\epsilon_s})^{-1}$, and angular momentum~$J_{\rm{EFS}} \propto \epsilon_s^{-1}$. We therefore focus on dimensionless quantities, e.g. compactness, when placing constraints using \eventname. 

In the bottom panel of Fig.~\ref{fig:solitonic-axionic-fluid}, we illustrate families of solutions at different axis ratios, where smaller $r$ typically results in more rapidly spinning and compact stars. We find solutions with spins compatible with \eventname to have $r \gtrsim 0.55$; note that we exclude unstable configurations beyond the maximum-mass turning point from our analysis. Models compatible with \eventname therefore enable us to place a lower limit on the allowed compactness of EFSs of $C \gtrsim 0.24$ (using a sequence of more oblate EFSs corresponding to the smallest $r$); however, this limit increases to $C \gtrsim 0.34$ for more spherical stars ($r \sim 0.7$).

\prlsec{Other ECOs} 
SIQMs have also been computed for a range of other ECO models that we now discuss here. For these models we are unable to place any meaningful constraints using the primary of \eventname due to either their incompatible spin predictions, or their SIQM values lying close to the Kerr value.

For example, AdS black shells and their corresponding SIQMs have been computed to sixth order in angular momentum~\cite{Danielsson:2023onu}. The configurations are considered reliable up to $\chi \leq 0.45$, which lies below \eventname spin measurement. 
Likewise, slowly rotating solutions of thin-shell GSs (with de-Sitter cores) have been constructed in Refs.~\cite{Uchikata:2021jmy, Uchikata2016}.
In contrast to AdS shells, these GSs were found to encompass spins beyond the Kerr range. 
A comprehensive understanding of the stability of GSs and AdS black shells further remains an open question~\cite{Danielsson:2021ykm,Danielsson:2021ruf,Uchikata2016,Horvat:2011ar}. However, at least for GSs, it was argued that membranes, with density and pressure having the same sign and obeying the weak energy condition, are unstable against high frequency oscillations~\cite{Yang:2022gic}. 

Scalarized BH solutions, arising in modifications to Einstein gravity
that add couplings between a new scalar degree of freedom and curvature invariants~\cite{Herdeiro:2015waa}, also predict SIQMs different
to Kerr BHs. However, within the range of coupling strengths where these solutions are known to exist, the differences in $\kappa$ from the Kerr value tend to be smaller than the range ruled out by \eventname -- on the order of $\lesssim 10\%$  within the relevant range, e.g., for Einstein-scalar-Gauss-Bonnet~\cite{Kleihaus:2015aje} and dynamical Chern-Simons gravity~\cite{Yagi:2012ya}. 

Finally, Proca stars (PSs)---vector analogues of BSs~\cite{Brito:2015pxa} ---are natural extensions to the cases considered above. Focusing on solutions with no self-interactions~\cite{Clough:2022ygm}, we compute SIQMs for rotating $m=1$ models using the methods outlined in Ref.~\cite{Siemonsen:2023hko}. We find no models satisfying the $\chi < 1$ bound. Further, all of the models constructed in the sequence (even less compact ones) have $\kappa \sim \mathcal{O}(1)$. PSs are therefore a special case where the SIQM is 
weakly correlated
with compactness.

\prlsec{Discussion} 
By employing the remarkably precise measurement of the SIQM, we placed constraints on an ECO being the primary in the \eventname merger event. Our analysis uncovered that BSs with repulsive potential can be completely ruled out, whilst highly compact ($C \sim 0.3$) solitonic and axionic stars may still be viable exotic contenders.
We additionally ruled out exotic fluid stars with the EOS at the causal limit with compactnesses of $C \lesssim 0.24$.
In general, over the models considered here, we found that all compact objects still permitted by
\eventname had compactnesses above this limit.  
Finally, we revisited theoretical predictions for SIQMs for GSs, AdS black shells, Proca stars, and dynamically scalarized BHs; however our constraints for these models were less conclusive.

Echoing Heisenberg~\cite{Heisenberg}—\textit{“what we observe is not nature itself, but nature exposed to our method of questioning”}—our results should be interpreted with care. Our analysis tests for one specific \textit{finite-size} effect relating to SIQMs, and by construction does not account for other possible effects, such as tidal deformability~\cite{Noah2017, Johnson-Mcdaniel:2018cdu,Cardoso:2017cfl} or BS-specific short-range interactions, that are known to play an important role in binary BS dynamics~\cite{Siemonsen:2023hko,Evstafyeva:2022bpr,Evstafyeva:2024qvp}. Furthermore, our waveform does not incorporate superextremal spins (possible in most rotating BS and GS models).
In the future, our analysis should include these effects, as well as higher PN orders~\cite{Mishra:2016whh} and higher multipole moments -- such as the spin-induced octupole -- which altogether could offer additional insights into the current constraints. 

A complete program for probing the nature of compact objects through inspiral, merger, and ringdown would ideally require (i) numerical and phenomenological construction of ECO template banks, (ii) cross-verification of consistency across various tests of general relativity and the standard model~\cite{Johnson-McDaniel:2021yge}, (iii) simultaneous probing of the true nature of compact objects using population inference~\cite{Magee:2023muf}, and/or (iv) complementary findings from other observational channels. 
In particular, item (i) can be mainly driven by phenomenological modeling, and supplemented by numerics in some parts of the parameter space. For example, in analogy to NSs~\cite{Yagi:2013awa}, (quasi)-universal relations for ECOs may provide an efficient means to constrain quantities that are otherwise not directly observable.

\eventname marks a crucial stepping stone towards probing the nature of compact objects and scrutinizing models of ECOs. Looking ahead, future GW observations and detectors promise exciting dialogues between theory and experiment, enabling increasingly stringent tests of the Kerr nature of compact objects. 

{\it Data availability:} Our boson-star data and their spin-induced quadrupole moments are available at~\cite{bs_data}. 

{\it Acknowledgments:} AV thanks Gonzalo Alonso-Alvarez, Reed Essick, and Gibwa Musoke for discussions.  We thank Elisa Maggio for carefully reading the manuscript and providing us with useful comments and suggestions. Authors thank K. G. Arun and Nathan Johnson Mc-Daniel for useful discussions at various stages of the project. Krishnendu thanks Patricia Schmidt and Frank Ohme for discussions. This document has LIGO preprint number {\tt P2500682}. 
AV acknowledges support from the Natural Sciences and Engineering Research Council of
Canada (NSERC) (funding reference number 568580). Krishnendu is supported by STFC grant ST/Y00423X/1. W.E. acknowledges support from a Natural Sciences and Engineering Research
Council of Canada Discovery Grant and
an Ontario Ministry of Colleges and Universities Early Researcher Award.
This research was
supported in part by Perimeter Institute for Theoretical Physics. Research at
Perimeter Institute is supported in part by the Government of Canada through
the Department of Innovation, Science and Economic Development and by the
Province of Ontario through the Ministry of Colleges and Universities. AG is supported by National Science Foundation grants PHY-2308887 and CAREER-2440327. P. D. R. also acknowledges support from the National Science Foundation (NSF) via NSF Award No. PHY-2409372. MS acknowledge the support from Weinberg Institute for Theoretical Physics at University of Texas at Austin. The authors are grateful
for computational resources provided by the LIGO Laboratory and supported by National Science Foundation Grants PHY0757058 and PHY-0823459 and the Symmetry cluster at Perimeter Institute. This research has made use of data or software obtained from the Gravitational Wave Open Science Center (https://www.gwosc.org), a service of the LIGO
Scientific Collaboration, the Virgo Collaboration, and KAGRA. This material is based upon work supported by NSF's LIGO Laboratory which is a major facility fully funded by the National Science Foundation.
\clearpage
\onecolumngrid
\setcounter{equation}{0}
\setcounter{figure}{0}
\setcounter{table}{0}
\makeatletter
\renewcommand{\theequation}{S\arabic{equation}}
\renewcommand{\thefigure}{S\arabic{figure}}
\renewcommand{\thetable}{S\arabic{table}}

{\centering
\begin{center}{\large Supplementary Material\\[3ex] \textbf{\fulltitle}}\end{center}
}
\section{Computation of spin-induced quadrupole moments}

\subsection{Overview of the formalism}

In this section, we outline the procedure used to extract spin-induced quadrupole moments (SIQMs) from numerically computed spacetimes, following the prescription of Refs.~\cite{Pappas:2012ns,Pappas:2012qg}. We begin with the line element for a stationary and axisymmetric spacetime in quasi-isotropic coordinates, adopting the notation of Butterworth and Ipser~\cite{Butterworth:1976},
\begin{equation} \label{eq:quasi_metric}
    \du s^2 = -e^{2{\nu}}\du t^2 + r^2 \mathrm{sin}^2 \theta B^2 e^{-2\nu} (\du \phi - \omega \du t)^2 + e^{2(\zeta - \nu)} (\du r^2 + r^2 \du \theta^2).
\end{equation}
Above, the quantities $\nu$, $B$, $\omega$, and $\zeta$ denote the metric functions of $(r, \theta)$. The asymptotic behaviour of Eq.~\eqref{eq:quasi_metric} has been derived in Ref.~\cite{Butterworth:1976} [see their Eqs.~(12a)-(12c)]; below we inlcude the first few leading terms
\begin{align} 
    \nu &\sim \left(-\frac{M}{r} + \frac{B_0 M}{3r^2} + ... \right) + \left(\frac{\nu_2}{r^3} + ... \right)P_2 + ..., \label{eq:nu_expansion} \\
    \omega &\sim \left(\frac{2J}{r^3} - \frac{6JM}{r^4} + ... \right) \frac{\du P_{1}(\mathrm{cos}\theta)}{\du \mathrm{cos} \theta} + \left(\frac{\omega_2}{r^5} + ... \right)\frac{\du P_{3}(\mathrm{cos}\theta)}{\du \mathrm{cos} \theta} +... \\
    B &\sim \sqrt{\frac{\pi}{2}} \left(1 + \frac{B_0}{r^2} \right) T_{0}^{1/2} + ..., \label{eq:B_expansion}
\end{align}
where $P_l(\mathrm{cos}\theta)$ denote Legendre polynomials, $T_l^{1/2}$ Gegenbauer polynomials (see Eq.(10a) of Ref.~\cite{Butterworth:1976}), $M$ and $J$ the first two multipole moments of the spacetime (i.e.~mass-energy and angular momentum). As first pointed out in Ref.~\cite{Pappas:2012ns}, care has to be taken when relating the asymptotic behavior of the metric coefficients of numerical spacetimes to the multipole moments of Hansen-Geroch~\cite{Geroch:1970cd,Hansen:1974zz}. Using Ryan's coordinate-independent method of identifying the multipole moments~\cite{Ryan:1995wh}, one finds that the quadrupole moment $M_2$ is given by~\cite{Pappas:2012ns}
\begin{equation} \label{eq:m2}
    M_2 = -\nu_2 - \frac{1}{3} M^3 - \frac{4}{3} B_0 M.
\end{equation}
The \textit{reduced} quadrupole is therefore
\begin{equation}
    \kappa_2 = - \frac{M_2}{\chi^2 M^3},
\end{equation} 
where $\chi = J/M^2$ is the dimensionless spin. In this work we compute the multipole moments using \eqref{eq:m2}; our methods in extracting the $\nu_2$ and $B_0$ coefficients differ across different ECOs we consider. We therefore discuss their details separately.
\subsection{Boson stars}
We numerically compute boson star solutions using the relaxation algorithm described in Ref.~\cite{Siemonsen:2020hcg}. We compute most of our spinning BS solutions using the resolution of $N_{\tilde{r}} \times N_{\theta} = 500 \times 50$, where $\tilde{r} = r/(1+r)$ denotes the compactified coordinate. Although for highly compact stars, we increase it even further to $N_{\tilde{r}} \times N_{\theta} = 500 \times 100$ and sometimes even $N_{\tilde{r}} \times N_{\theta} = 1000 \times 100$. The numerical error incurred in the computation of the reduced quadrupole at these resolutions varies between $0.3-1.1\%$, depending on the compactness of the model in question. 

The metric of the BS spacetime in our case is assumed to be
\begin{equation} \label{eq:metric_boson}
    \du s^2 = - f \du t^2 + lf^{-1} \{g \left(\du r^2 + r^2 \du \theta^2 \right) + r^2 \mathrm{sin}^2 \theta \left(\du \phi - \Omega r^{-1} \du t \right)^2 \},
\end{equation}
where $l, f$ and $\Omega$ are functions of $(r, \theta)$. Our form is similar to that of Eq.~\eqref{eq:quasi_metric}, however it requires some metric reparametrizations to be matched exactly, e.g.,~$\nu = \mathrm{ln}(f)/2$ and $B = \sqrt{l}$. The required ingredients for the computation of $M_2$ can then be easily extracted from Eqs.~\eqref{eq:nu_expansion}-\eqref{eq:B_expansion} using orthogonality conditions for Legendre and Gegenbauer polynomials
\begin{align} 
    \nu_0 &= -M = \frac{1}{2} \lim_{r \to \infty} r \int_{0}^{\pi} \nu(r, \theta) \mathrm{sin}\theta d\theta, \label{eq:nu0_coeff} \\
    \nu_2 &= \frac{5}{2} \lim_{r \to \infty} r^3 \int_{0}^{\pi} \nu(r, \theta) \frac{1}{2} (3 \mathrm{cos}^2\theta - 1)  \mathrm{sin}\theta d\theta, \label{eq:nu2_coeff} \\
    B_0 &= \lim_{r \to \infty} r^2 \int_{0}^{\pi} \sqrt{\frac{2}{\pi}} \left[B(r,\theta) - 1 \right] \mathrm{sin}^2\theta d\theta. \label{eq:b0_coeff}
\end{align}
Our expressions in Eqs.~\eqref{eq:nu0_coeff}-\eqref{eq:b0_coeff} are also in agreement with Ref.~\cite{Adam:2022nlq}. 

\subsection{Exotic fluid stars}

As described in the main text, our EFS models are computed using the \textsc{rns} package~\cite{Nozawa:1998ak,Stergioulas:1994ea,Cook1992:amns}. The metric takes the form of
\begin{equation}
\du s^2 = -e^{\gamma+\rho}\du t^2 + r^2 \mathrm{sin}^2 \theta e^{\gamma-\rho} (\du \phi - \omega \du t)^2 + e^{2 \alpha} (\du r^2 + r^2 \du \theta^2),    
\end{equation}
where again $\gamma, \rho, \omega$ and $\alpha$ are functions of $(r, \theta)$ and one requires further metric reparametrizations to match it to the metric form of Butterworth and Ipser: $B = e^{\gamma}$ and $\nu = (\gamma + \rho)/2$. The SIQMs are extracted using Eq.~\eqref{eq:m2}, as for BSs. However, here we utilize the integral expressions using the methods of Komatsu, Eriguchi and Hachisu~\cite{Komatsu:1989ikr} from Ref.~\cite{Cook1992:amns} [see their Eqs.~(27)-(29)] to compute $\nu_2$ and $B_0$ coefficients. We use resolution of $481\times301$ for radial and angular grid-points and estimate the reduced quadrupole moment errors to be $\sim 1\%$.   

\section{GW measurement of the SIQM parameter}
\label{sec:appSIQM}
\subsection{Waveform model and Bayesian inference }

As already discussed in the main text, we perform parameter estimation on \eventname\ using the \textsc{IMRPhenomXPHM} waveform model~\cite{Pratten:2020ceb, Divyajyoti:2023izl}, which includes SIQM corrections in the inspiral phase at 2PN and 3PN orders. Explicit expressions for these terms are given in Eqs.~(1.5a) and (1.5b) of the Supplementary Material of Ref.~\cite{Krishnendu:2017shb}.

The marginalized posterior distribution of $\kappa$ is obtained using Bayes’ theorem,
\begin{equation}
p(\kappa|d,\mathcal{H}) = \int p(\vec{\theta}|d,\mathcal{H})\, d\vec{\theta},
\label{eq:posterior-dks}
\end{equation}
where
\begin{equation}
p(\vec{\theta}|d,\mathcal{H}) = \frac{\pi(\vec{\theta}|\mathcal{H})\, \mathcal{L}(d|\vec{\theta},\mathcal{H})}{Z(\mathcal{H})}.
\label{eq:posterior}
\end{equation}
Here, $\vec{\theta}$ denotes the full parameter set (binary black hole parameters and $\kappa_i$, $i \in [1,2]$), $\mathcal{L}(d|\vec{\theta},\mathcal{H})$ the likelihood, $\pi(\vec{\theta})$ the prior, and $Z$ the evidence.

\begin{figure}[!htbp]
    \includegraphics[width=3.5 in]{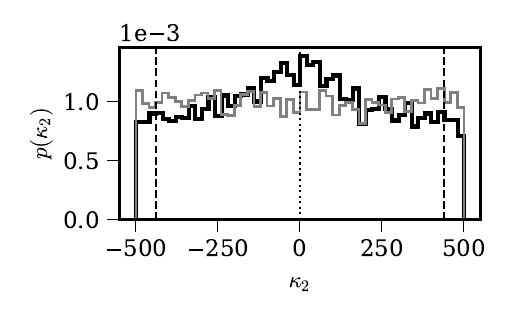}
    \caption{The one-dimensional posterior distribution of $\kappa_2$ obtained from \eventname along with the prior distribution in grey.}
    \label{fig:K2_1D}
\end{figure}

The likelihood function $\mathcal{L}(d|\vec{\theta},\mathcal{H})$ is constructed assuming stationary Gaussian noise and a model consistent with the Kerr black hole hypothesis $\mathcal{H}$.
The likelihood function is evaluated between a minimum frequency $f_\mathrm{min}$ (typically $20\,\mathrm{Hz}$ for current-generation LIGO and Virgo detectors) and a maximum cutoff frequency $f_\mathrm{cut}$~\cite{Thrane:2018qnx,VeitchVecchio09,lalsuite,bilby_paper}.
Posterior distributions are sampled using the \textsc{Dynesty} nested sampler~\cite{speagle2020dynesty,dynesty} within the \textsc{Bilby} framework~\cite{Ashton:2018jfp,Romero-Shaw:2020owr,pbilby_paper}.
We assume uniform priors on $\kappa_1$ and $\kappa_2$ over the range $[-500,500]$~\cite{gw241011}.
In our notation, $\kappa_1$ ($\kappa_2$) corresponds to the primary (secondary) component of the binary, with $m_1 > m_2$.
Alongside $\kappa_1$ (Fig.~\ref{fig:K1_1D}), we simultaneously infer $\kappa_2$, shown in Fig.~\ref{fig:K2_1D}.
Since the secondary in \eventname\ is slowly spinning, the $\kappa_2$ posterior is uninformative, and we cannot draw conclusions about its nature.

\subsection{Justification for the choice of the upper cut-off frequency, $f_{\rm {cut}}$}
\label{subsec:fmax}

\begin{figure}[!htbp]
    \includegraphics[width=3.5 in]{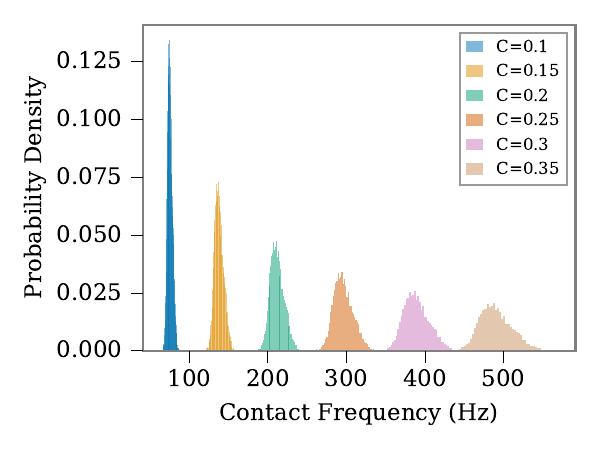}
    \caption{The approximate contact frequency $f_{\rm contact}$ for different compactnesses.}
    \label{fig:fmax_supl_compactness}
\end{figure}

\begin{figure}[!htbp]
    \includegraphics[width=4.5 in]{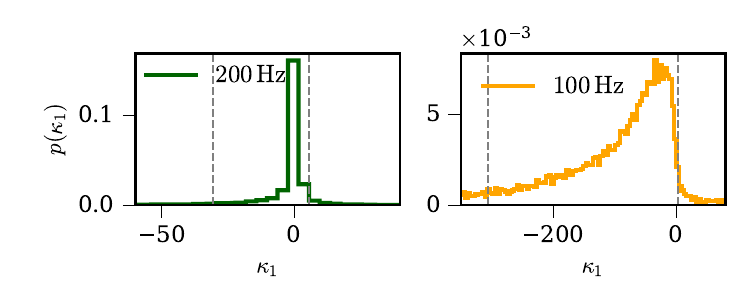}
    \caption{The one-dimensional posterior distribution of $\kappa_1$ obtained from \eventname and estimated from $f_{\rm cut}=200$ (left panel) and  $100 \, \rm Hz$ (right panel).}
    \label{fig:fmax_supl}
\end{figure} 

The SIQM effects are implemented only in the inspiral portion of the waveform model. However, the likelihood evaluation in Fig.~\ref{fig:K1_1D} of the main text extends up to $f_{\rm cut}$, thereby including the post-inspiral regime. From a null-test perspective—where no deviations from the Kerr black hole hypothesis are assumed—this treatment is appropriate, as it allows the full signal to be used in a model-agnostic consistency test. The resulting $\kappa$ constraints are primarily informed by the inspiral phase, where the Kerr description of the binary components is expected to hold. As our waveform is parametrized in terms of explicit $\kappa$ dependence, any potential deviation from the Kerr hypothesis would manifest as a shift of the posterior peak away from unity. Nevertheless, when translating these results to specific exotic compact object models, it is important to assess whether including the post-inspiral data introduces systematic biases.

To quantify these systematic effects, if present, we evaluate posterior distributions using different choices of the upper frequency cutoff in the likelihood computation. Specifically, we consider truncations at $f_{\rm cut}=529\,\mathrm{Hz}$, $489\,\mathrm{Hz}$, $439\,\mathrm{Hz}$,  $400\,\mathrm{Hz}$, $300\,\mathrm{Hz}$, $200\,\mathrm{Hz}$, and $100\,\mathrm{Hz}$, as well as the full frequency range used in the generic analysis ($f_{\rm cut}=1024\,\mathrm{Hz}$). This procedure allows us to test the robustness of the inferred $\kappa$ constraints and identify potential biases introduced by including signal beyond the inspiral regime. 
The higher value of $f_{\rm{cut}} = 489\,\rm{Hz}$ is motivated by the inner most stable circular (ISCO) frequency of a Kerr BH~\cite{Favata2010, Krishnendu:2018nqa}, derived from the maximum likelihood estimates of the \eventname masses and spins, while lower truncation frequencies are motivated by the expected contact frequencies $f_{\rm contact}$, at which the surfaces of two compact objects would come into contact. This value typically depends on the mass and compactness, and it can be \textit{approximated} by $f_{\rm{contact}} = (c^3/2G) (C^{3/2} / m_{\rm{ECO}})$, where $m_{\rm{ECO}}$ is the mass of the ECO and $C$ is its compactness~\cite{Ghosh:2025wex}. Using the posterior of the primary mass of \eventname, in Fig.~\ref{fig:fmax_supl_compactness}, we illustrate the contact frequency estimates for $C \in [0.1-0.35]$. The contact frequency spans a wide range, from $\sim 89\,\mathrm{Hz}$, insufficient for a meaningful constraint, to $\sim 500\,\rm{Hz}$, comparable to the Kerr black hole ISCO frequency. 

\begin{figure}[!htbp]
    \includegraphics[width=5.5 in]{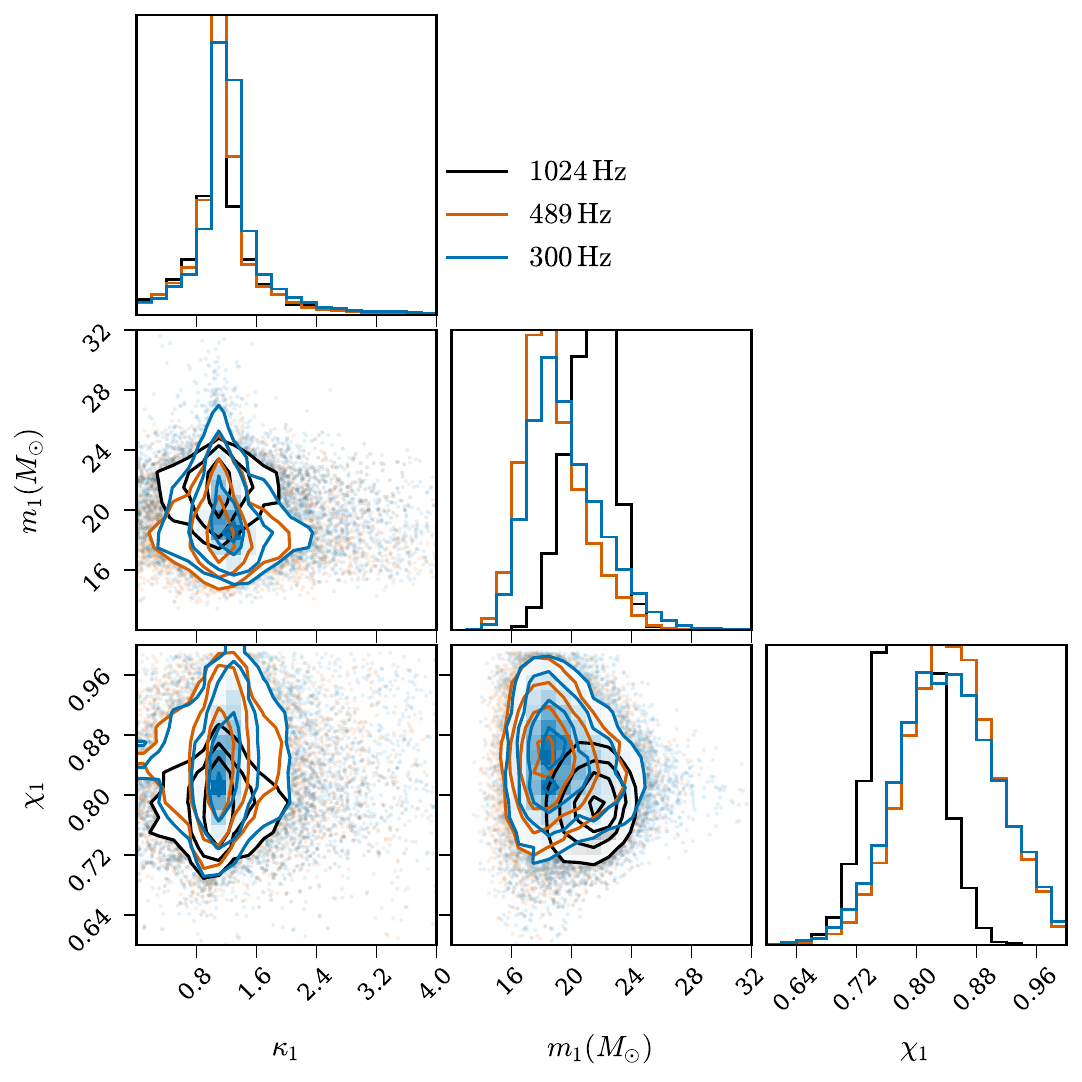}
    \caption{The correlation between $\kappa$, with primary mass and spin for different $f_{\rm cut}$ values.} 
    \label{fig:fmax_supl_corner}
\end{figure}

The black histogram in Fig.~\ref{fig:K1_1D} of the main text corresponds to the analysis using the full frequency range ($f_{\rm cut}=1024$ Hz), while the results for the other cutoff frequencies are shown as distinct colored histograms. For the lower cutoff frequencies, e.g., $f_{\rm cut}=100$–$200\,\mathrm{Hz}$, the posterior distributions are noticeably broader, making direct comparison in Fig.~\ref{fig:K1_1D} less clear; these low-frequency cases are therefore displayed separately in Fig.~\ref{fig:fmax_supl}. Overall, the inferred parameters remain broadly consistent across all frequency choices (cf. Fig.~\ref{fig:fmax_supl_corner}, where representative posteriors are overplotted). However, the $100\,\mathrm{Hz}$ case, where the limited signal bandwidth leaves the binary parameters largely unconstrained, yields comparatively less tight $\kappa_i$ estimates.
As the cutoff frequency decreases, the $\kappa$ posterior distribution becomes noticeably broader on the negative side. This asymmetry arises because the reduced signal strength weakens constraints on the binary  masses and spins, thereby enhancing the mass–spin–$\kappa$ degeneracy. For a positively spinning primary, this naturally leads to an extended, weakly constrained region toward negative $\kappa$ values~\cite{Krishnendu:2019tjp}.

\begin{figure}
    \centering
    \includegraphics{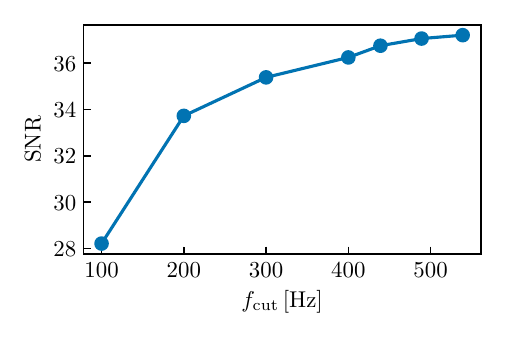}
    \caption{The SNR accumulated in the signal as a function of $f_\mathrm{cut}$. }
    \label{fig:supl-k_correlation}
\end{figure}

 Finally, we also plot the the coherent network signal-to-noise ratio (SNR) as a function of $f_\mathrm{cut}$ in Fig.~\ref{fig:supl-k_correlation}. We see that there is consistent build-up of SNR far past the nominal values of $f_\mathrm{cut}$ of $200$-$300\, \mathrm{Hz}$, suggesting that the objects do not come into contact at these frequencies. 

\clearpage
\twocolumngrid
\bibliographystyle{apsrev4-1}
\bibliography{refs}
\end{document}